\RequirePackage{lineno}
\documentclass[aps,prl,nolongbibliography,twocolumn,floatfix,superscriptaddress]{revtex4-1}

\usepackage{amsmath}
\usepackage{amssymb}
\usepackage{amsthm}
\usepackage{mathrsfs}
\usepackage{mathtools}
\usepackage{graphicx}
\usepackage{setspace}
\usepackage{float}
\usepackage{xcolor}
\graphicspath{{figures/}} % Directory in which figures are stored

\begin{document}

\title{A generic coupling between internal states and activity leads to 
activation fronts and criticality in active systems}

%% \title{Activation waves, fluctuations and criticality as direct consequence of coupling two internal states with active motion}

%% Author 1
\author{Hadrien-Matthieu Gascuel}
\email{equally contributed}
\affiliation{CNRS, Centre de Recherches sur la Cognition Animale, 118 route de Narbonne, F-31062 Toulouse Cedex 9, France.}
\affiliation{Universit{\'e} C{\^o}te d'Azur, Laboratoire J. A. Dieudonn{\'e}, UMR 7351 CNRS, Parc Valrose, F-06108 Nice Cedex 02, France.}

\author{Parisa Rahmani}
\email{equally contributed}
\affiliation{Laboratoire de Pysique Th{\'e}orique et Mod{\'e}lisation, UMR 8089, CY Cergy Paris Universit{\'e}, 95302 Cergy-Pontoise, France}

%% Author 2
\author{Richard Bon}
\affiliation{CNRS, Centre de Recherches sur la Cognition Animale, 118 route de Narbonne, F-31062 Toulouse Cedex 9, France.}

%% Author 3
\author{Fernando Peruani}
\email{fernando.peruani@cyu.fr}
\affiliation{Universit{\'e} C{\^o}te d'Azur, Laboratoire J. A. Dieudonn{\'e}, UMR 7351 CNRS, Parc Valrose, F-06108 Nice Cedex 02, France.}
\affiliation{Laboratoire de Pysique Th{\'e}orique et Mod{\'e}lisation, UMR 8089, CY Cergy Paris Universit{\'e}, 95302 Cergy-Pontoise, France}

\date{\today}

%%
%% Paper starts here
%%

%% Uncomment the line below to double-space the document
%% \doublespacing

\begin{abstract}
To understand the onset of collective motion, we investigate active systems where particles switch on and off their self-propulsion. 
We prove 
that even when the only possible transition is off$\to$on, an active 2-state system behaves as an effective 3-state system 
that exhibits a sharp phase transition in 1D, and critical behavior in 2D, with scale-invariant activity avalanches. 
The obtained results show how criticality can naturally emerge in active systems, providing insight into the way collectives distribute, process, and respond to local environmental cues.
\end{abstract}

%\begin{abstract}
%We investigate a generic coupling between the particle's  internal state and self-propulsion to study the onset of collective motion. Our analysis reveals such a coupling renders an otherwise non-critical 2-state system, into an effective 3-state system able to display scale-invariant activity avalanches, proving the existence of an intimate connection with critical phenomena. Furthermore, we identify three distinct propagating front regimes, including a selfish regime. The obtained results provide insight into the way collectives distribute, process, and respond to local environmental cues.
%\end{abstract}

\maketitle

%% \linenumbers
%\section{Introduction}
Collective motion is observed in a large variety of biological systems; fish schools~\cite{lopez2012behavioural}, bird flocks~\cite{ballerini2008interaction}, and ungulate herds~\cite{ginelli2015intermittent, gomez2022intermittent} are a few of the countless existing examples~\cite{krause2002living,vicsek2012collective}. 
Despite the fact that collective motion is in general not a continuous process~\cite{o1990search,kramer2001behavioral}, and animal groups display repeated transitions from static to moving phases~\cite{gomez2022intermittent} -- with the former associated with e.g. resting or feeding phases -- most experimental and theoretical studies have focused on the characterization and modeling of groups of constantly moving units~\cite{lopez2012behavioural,ballerini2008interaction,vicsek2012collective}. 
Moreover, transitions from static to moving states have been recently shown to also occur in (subcritical) active colloidal systems~\cite{liu2021activity}, which proves that such transitions are  observed in both, living and non-living active matter. 

A prominent example of such transitions is the onset of collective motion from an initially polarized static group, which remains largely unexplored. One exception is a recent study on the initiation of a Marathon, where boundary displacements  set in motion the crowd~\cite{bain2019dynamic}. 
In animal groups, on the other hand, collective motion is often triggered by the behavioral shift, from static to moving, of one single individual~\cite{pillot2011scalable,toulet2015imitation,ginelli2015intermittent,strandburg2015shared}.
The behavioral shift of this first individual can be the result of, for instance, the decision of the individual to search for a new feeding area, or a  reaction to a predator attack~\cite{kramer2001behavioral}. 
%
% And thus, we can imagine that this first individual performs a slow displacement in the former scenario and a fast one in the latter.  
%
In general, it is expected that certain features of the behavioral shift of this first individual -- e.g. its velocity  --   encode information about the stimulus that triggered its behavioral change. 
%
% Thus, the statistical properties of collective motion initiation  are likely to provide insight into how collectives sense and distribute local environmental information. 

Understanding how information spreads in active and animal systems remains an open crucial question. It has been argued~\cite{mora2011biological,munoz2018colloquium,klamser2021collective} that  biological systems 
have to operate at criticality to ensure efficient responses and fast information propagation to  external perturbations.  
In the context of animal behavior, evidence of critical behavior was reported in bird flocks~\cite{ballerini2008interaction,cavagna2010scale,attanasi2014information}, fish schools~\cite{poel2022subcritical}, and sheep herds~\cite{ginelli2015intermittent}. 
Information propagation, on the other hand, has been mostly studied, not on active systems, but on static lattices and  networks~\cite{castellano2009statistical}. 
Except for the study of epidemic spreading in moving agent systems~\cite{peruani2008dynamics, peruani2019reaction, rodriguez2019particle, forgacs2022using, zhao2022contagion, PhysRevResearch.4.043160}, the role of agent motility on a propagation process, despite its relevance, is unknown. 
Moreover, in these models~\cite{peruani2008dynamics, peruani2019reaction, rodriguez2019particle, forgacs2022using, zhao2022contagion}, with the exception of~\cite{levis2020flocking, azais2018traveling, paoluzzi2020information}, active  agents do not exhibit a coupling between the  internal state dynamics and the equation of motion of the agents.  

\begin{figure}
\centering
\includegraphics[width=1.0\columnwidth]{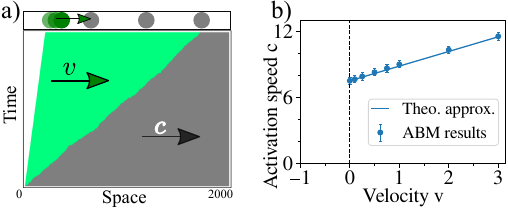}
\caption{Information propagation in a 1D system.   
 a) The kymograph  shows the evolution of the activation wave in simulation. Green corresponds to density of active and gray inactive agents. The leading edge moves at speed $c$, Eq.~(\ref{eq:InfoSpeed}). The upper rectangle illustrates the process with agents represented as disks. See video S1 in~\cite{SI}. 
b) Propagating front speed  $c$ vs $v$. 
Blue symbols correspond to simulations, solid line is Eq. (\ref{eq:InfoSpeed}).}
\label{fig:MassInfoSpread}
\end{figure}

Here, we fill this fundamental gap and investigate a generic coupling in which the agent's internal state controls the agent's active speed. Specifically, 
we consider systems in which the state of an agent $i$ is given by its position $x_i$ and its behavioral state $q_i$ that adopts one of the two possible values: I (inactive) or A (active). 
The generic feedback mechanism between the internal state $q_i$ and the spatial dynamics of the agent is given by: 
\begin{equation}\label{eq:motion}
 \dot{x}_i = \text{V}[q_i] =
   \begin{cases} 
 v & \text{if } q_i = \text{A} \\
   0       & \text{if } q_i = \text{I}
  \end{cases} \, ,
\end{equation}
where the dot denotes the temporal derivative and $v$ the active velocity.  
Initially, individuals are equally spaced a distance $\Delta_0=|x_{i+1}(t=0)- x_{i}(t=0)|$, where, in 1D, $i=1$ and $i=N$ are the leftmost and rightmost agents, respectively,
and all individuals are in state I, except for one agent (usually $i=1$) that is in state A (Fig.\ref{fig:MassInfoSpread}(a) upper panel). This condition corresponds to an initially  polarized static group of agents and one  initiator. 
%
% For simplicity, here we focus only on one  behavioral transition.  
Given an agent in state I and another one in state A, separated by a distance $|\Delta x|$, we consider only one possible transition: 
\begin{equation}\label{eq:reaction}
\text{I}  + \text{A} \underset{\gamma(|\Delta x|) }{\longrightarrow} 2\text{A} \, ,
\end{equation}
where  $\gamma(|\Delta x|)=\alpha\, K(|\Delta x|/d)$ denotes the transition rate, with $\alpha$ a constant rate, $d$ a characteristic length, and $K(u)=\text{e}^{-|u|}$. Note that we have  tested  $K(u)=\text{e}^{-u^2}$, step-functions, and other functional forms for $K(u)$, obtaining (qualitatively) the same results~\cite{SI}. 
%$K(u)=H(u-1)$ with $H(\cdot)$ a Heaviside step function 
%
For $v=0$, the model describes the propagation of state A on a lattice such that   
for $t\to \infty$, all  individuals end up into state $A$.
As explained below, for $v>0$ the system exhibits different propagation regimes that are controlled by the active speed $v$. 
For $v<0$, the simple generic coupling given by Eq.~(\ref{eq:motion}) fundamentally changes the physics of the problem. 
The system behaves as an effective 3-state model that exhibits a sharp phase transition in 1D and critical behavior in 2D.  

We observe that the model can be interpreted as follows. The first individual that becomes active -- for instance, due to some external perturbation, e.g. the presence of a predator -- chooses a velocity $v$. All subsequent transitions I$\to$A  
 can be interpreted as a mimetic process: inactive agents as transition to active, adopt the velocity of the other
active agents, and thus, acquire the velocity of the first agent. 
% 
% Below, we explain how sensitive the propagation process is to this initial velocity $v$.

%
%
% Notice that in the thermodynamic limit there is no asymptotic propagation in one-dimensions for $v<0$.  
%\section{Information speed}

\begin{figure}
\centering
\includegraphics[width=1.0\columnwidth]{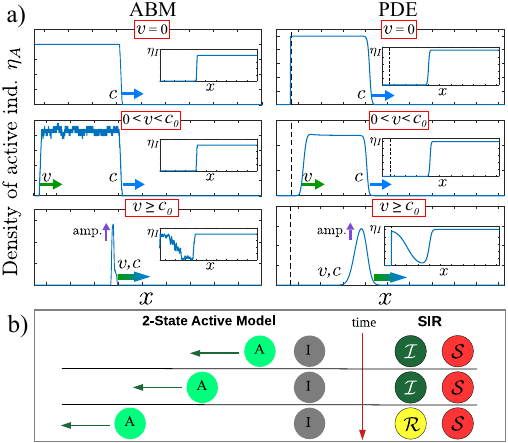}
\caption{Dynamic regimes.   
a) Density profile of active agents $\eta_A(x,t)$ at a fixed time $t$ in agent-based simulations (ABM) and in the numerical integration of Eq. (\ref{eq:macro}) (PDE).
Three different regimes are shown: the limit $v=0$, the slow velocity regime ($0<v<c_0$), and the HS or  selfish regime ($v\geq c_0$).   
The insets display the density of inactive agents $\eta_I(x,t)$ at time $t$.  
b) Mapping of our 2-state model with spatial coupling onto the 3-state SIR model (color code as in Fig. {\ref{fig:MassInfoSpread}}). 
}
\label{fig:densityProfile}
\end{figure}

{\it Propagating regimes and fronts.--}
To understand the presence of distinct propagating regimes and the relation between the propagating front speed  $c$ and the active agent velocity $v$, 
we derive the associated Kolmogorov forward equation~\cite{gardiner1985handbook} corresponding to Eq.~(\ref{eq:motion}) and transition (\ref{eq:reaction}), and obtain a hydrodynamic description of the model  in terms of a density field $\eta_A(x,t)=\langle \sum_i \delta(x-x_i(t)) \delta_{q_i(t),A} \rangle$ [$\eta_I(x,t)=\langle \sum_i \delta(x-x_i(t)) \delta_{q_i(t),I} \rangle$] of active [inactive] individuals at time $t$ in position $x$: 
 \begin{subequations}
 \label{eq:macro}
 \begin{align}
 \partial_t \eta_A(x,t) &= \, \Gamma(x,t)\ \eta_I(x,t) - v\ \partial_x \eta_A(x,t) \label{eq:macro1}\\
 \partial_t \eta_I(x,t) &= \text{-} \Gamma(x,t)\ \eta_I(x,t) \, , \label{eq:macro2}
 \end{align}
 \end{subequations}
 where $\Gamma(x,t)= \int_{\text{-} \infty}^{\infty} \eta_A(x',t) \gamma(|x'-x|) dx'$. 
 %
 % Note that the density of individuals, $\rho = \eta_A + \eta_I$, is such that $\partial_t \rho=-\partial_x J$, with $J=v\, \eta_A$ the agent mass flux.   
 %
For $v=0$ and after approximating  $\Gamma(x,t) \simeq d \alpha \left( a\eta_A + \frac{d^2b}{2} \partial^2_x \eta_A \right)$, where  $a=\int_{\text{-} \infty}^{\infty} K(|u|) du$, and $b=\int_{\text{-} \infty}^{\infty} u^2 K(|u|) du$ -- for details see~\cite{SI} --  
 it is possible to show that Eq.~(\ref{eq:macro}) exhibits propagating fronts that move at speed $c_0 \sim\alpha \Delta_0^{-1} \sqrt{2\,a\,b}$. 
In this limit, Eq.~(\ref{eq:macro}) exhibits a behavior  similar to the one of the Fisher-Kolmogorov-Petrovsky-Piskunov (FKPP) equation~\cite{van2003front}; Fig.~(\ref{fig:densityProfile}(a) upper panels) and videos S2 in~\cite{SI}. 
On the other hand,  for $v\neq0$, the behavior of Eq.~(\ref{eq:macro})  is fundamentally different from, and not reducible to the one of the FKPP equation.  
When the agent speed is such that $0\!<\! v \!<\!c_0$, the density profiles of the expanding active population (of a semi-infinite system of agents located initially at $x>0$)  can be approximated by a plateau, where the right (R) and left (L)  edges are given, respectively, by  $y_{R(L)}(x) = A_{R(L)} [1 \mp \tanh(x-x_{R(L)}(t))/\ell_{R(L)}]$ with 
 $A_{R(L)}$,  $\ell_{R(L)}$ constants. The position of the right and left edges are $x_{R}(t)\propto c t$ and $x_{L}(t)\propto v t$ and thus, advance  at speeds $c$ and $v$, respectively; Fig.~(\ref{fig:densityProfile}(a) middle panels), Fig. S1, and videos S3 in~\cite{SI}.
In agent-based model (ABM) simulations as well as by integrating Eq.~(\ref{eq:macro}), we find that the speed $c$ of the leading edge 
exhibits a generic linear dependency with the active agent velocity $v$: 
\begin{equation}
c(v) \simeq c_0+ m\, v \, , 
\label{eq:InfoSpeed}
\end{equation}
where, importantly, $m\neq1$. 
For an exponential function $K$ and in the explored range of $\Delta_0$, we obtain $m=\frac{4}{3}$ in both, simulations and from Eq.~(\ref{eq:macro}); Fig.~\ref{fig:MassInfoSpread} and Fig. S2 in~\cite{SI}. 
%
% {\color{red}Note that $c$ depends not only on the agent speed, i.e. $|v|$, but also on the moving direction, i.e. $\text{sgn}(v)$: the propagation is faster 
% when active agents move towards the group than when they do it away from the group (Fig.~\ref{fig:MassInfoSpread}(a)-\ref{fig:MassInfoSpread}and S???. Later we show that, for large system sizes in 1D, the propagation only occur for $v\geq0$.}

Another propagation regime -- which we refer to as high-speed (HS) (or ``selfish") regime -- is observed when $v\geq c_0$.
The initial active agent moves so fast that goes by inactive agents, which do not have the time 
to transition to the active state and are left behind. Eventually, the fast-moving active agent recruits  a second agent that starts running near the initial active agent,  which contributes to recruiting a third active agent, and so on, until a collective response emerges; see  Fig.~(\ref{fig:densityProfile}(a) lower panels and video S4. 
 This regime is also reproduced by Eq.~(\ref{eq:macro}), where we observe as in ABM simulations, that an initial small density profile of active agents grows in height and width over time, asymptotically converging to a propagating front (see videos S5~\cite{SI}).  
% ACA 
To discuss the relevance of these regimes, let us consider a hypothetical scenario in which the first active individual -- i.e. the left most individual -- reacts to the presence of a predator (located further to the left). 
If this individual chooses to move with speed  $v < c_0$, an activation wave will propagate through the system and all individuals will transition to the active state, 
and the group will move to the right away from the predator.  
Importantly, the structure of the group will remain the same, and consequently that the left most individual, and thus the closest to the predator, will always be the initiator. 
Hamilton selfish herd hypothesis~\cite{hamilton1971geometry}, however, suggests that 
individuals may try to protect themselves by moving to the inside of the group, leaving other agents exposed to the predator.  
Logically, the strategy requires individuals to modify the group structure. 
The initiator to stop being the left most individuals needs to move with a speed  $v > c_0$.
Only using such high speeds, the initiator will manage to penetrate inside the group and leave a number of agents between 
its position and the predator.
%
\iffalse
{\color{red} 
%This implies that if we imagine the activation of the first active individual results from a reaction to a predator attack, by moving with velocity $v\geq c_0$, this first active individual is able to penetrate inside the group and leave a number of agents between its position and the predator. 
On the other hand, if the initial active agent chooses to move with 
$v \ll c_0$, the activation propagates over all inactive agents, triggering a collective displacement of the group away from the predator, but where the closest individual to the predator is always the initial active agent.}  
\fi
%
%
%\section{Fluctuations}

{\it From two to effective three states.--} 
Fluctuations, which cannot be analyzed by the deterministic approach given by Eq.~(\ref{eq:macro}) -- play a fundamental role in the dynamics of the system [Eqs.~(\ref{eq:motion}) and~(\ref{eq:reaction})]. 
To understand such fluctuations, we start out by considering $v<0$ and a system composed of only two individuals, in states $A$ and $I$,  located at positions $x_A(t) = -|v|\, t - \Delta_0$ and $x_I(t) = 0$, respectively (see Fig.~\ref{fig:densityProfile}(b)).  
We are interested in knowing the probability $p_I(t)$ that the individual, initially inactive, remains in state $I$ at time $t$.  
This probability obeys $\partial_t p_I = -\alpha\, K(|x_A(t)|/d) p_I$, with initial condition $p_I(t=0)=1$.   
The solution of this equation, with $K(u)=\exp(-|u|)$, reads: 
%% $p_I(t)= \exp{\left[ - \int_0^t \alpha\, K(|x_A(t')|/d) dt'\right]}$. 
%
\begin{equation}
\label{eq:inactive}
 p_{I}(t)=\exp{\!\left[ - \frac{\alpha \, \,e^{-\Delta_0/d}}{\left( |v|/d \right)}\left( 1- e^{-\frac{|v|}{d}t} \right) \right]}\, .
\end{equation}
For $v\to0$, it is evident that $p_I(t\to\infty) \to 0$, meaning that certainly, the activation of the initially inactive agent has occurred. 
On the other hand, for a non-zero $v$,  $p_I(t\to\infty)>0$. 
This implies that with probability $p_T = 1-p_I(t\to\infty)<1$  the activation of the initially inactive agent has occurred. 

In the following, we prove that the problem can be mapped onto the susceptible-infected-recovered (SIR) model or Forest Fire model~\cite{kermack1927contribution,pastor2015epidemic,bak1990forest}. The SIR model  is defined by the reactions  $\mathcal{S}+\mathcal{I}\overset{\sigma}{\rightarrow} 2\mathcal{I}$ and $\mathcal{I}\overset{\rho}{\rightarrow} \mathcal{R}$.  
In a system with only two agents, initially in states $\mathcal{S}$ and $\mathcal{I}$, 
the probability of finding the initially susceptible agent in state $\mathcal{S}$ at time $t$ reads: 
$p_{\mathcal{S}}(t)=\exp{\big[\text{-}\frac{\sigma}{\rho} \big(1-e^{\text{-}\rho t}\big)\big]}$, 
and thus $p_{\mathcal{S}}(t\!\to\!\infty)=\exp{\big[\text{-}\frac{\sigma}{\rho} \big]}$. 
By taking $\rho \leftarrow |v|/d$ and $\sigma \leftarrow \alpha e^{-\Delta_0/d}$, it becomes evident that  $p_{\mathcal{S}}(t)$ is identical to Eq.~(\ref{eq:inactive}). 
Let us know consider a semi-infinite lattice, initially in the configuration $\mathcal{I}\mathcal{S}\mathcal{S}\mathcal{S}\hdots$, with interaction to nearest neighbors. 
If state $\mathcal{R}$ is not present, all agents end up being infected. 
Otherwise, if at some point an agent in state $\mathcal{I}$ transitions to $\mathcal{R}$ before passing over the disease to its neighbor, the infection cascade is interrupted (see Fig.~\ref{fig:densityProfile}(b)). 
The probability of observing $n$ lattice sites in state $\mathcal{R}$ for $t\!\to\!\infty$ is 
$(1-p_{\mathcal{S}}(t\!\to\!\infty))^n p_{\mathcal{S}}(t\!\to\!\infty)$. This implies that outbreak sizes ($n$) are exponentially distributed and the average size remains finite. 
Similarly in our 1D active model, the probability $P(s)$ of observing  $s = n_A(t\to\infty)$ active agents at $t\to\infty$, involves $s$ successive activation transmissions followed by a transmission failure, which   can be expressed as  $P(s)\! =\! \left[p_T\right]^s \!p_I(t\to\infty) \simeq \! e^{-s p_I(t\!\rightarrow\!\infty)} p_I(t\!\rightarrow\!\infty)$, which leads to  
 $\langle s\rangle \sim 1/ p_I(t\rightarrow\infty)$. 
As in the SIR model in 1D, activation cascades are exponentially distributed and the average number of active individuals at $t\to\infty$ -- $\langle s \rangle \simeq \int_0^{\infty} s\, p(s) ds$  -- remains finite. All this demonstrates the existence of a mapping between our 1D active motion with $v<0$ and the SIR model in 1D.   
Defining as order parameter $\phi =\langle s \rangle/N$, it is evident that in 1D for $v<0$ and in the thermodynamic limit, $\phi \to 0$. 
For $v=0$, on the other hand, we have seen that $p_I(t\!\to\!\infty)=0$ and  
$p_T=1$, and thus in the thermodynamic limit $\langle s \rangle \to \infty$ and $\phi\to 1$. 
The same applies to $v>0$; left most agent in state A has always a neighbor at a distance less or equal to $\Delta_0$. In summary, in 1D, $\phi=0$ for $v<v_c$ and $\phi=1$ for $v\geq v_c$ with $v_c=0$.  
Importantly, all of this proves that our active system in 1D, where active particles move, behaves as a lattice model with effective three states.

\begin{figure}
\centering
\includegraphics[width=1.0\columnwidth]{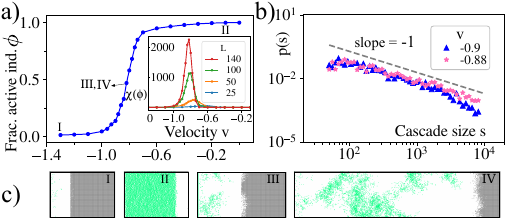}

\caption{Criticality in the 2D system.   
a) Order parameter $\phi=\langle s \rangle/N$ vs $v$. Inset shows susceptibility $\chi(\phi)$ vs $v$ for different system sizes $N=L^2$. 
b) Probability $p(s)$ of observing a cascade of size $s=n_A(t\to\infty)$ for $v \sim v_c$. 
For the result with one initial activator, see~\cite{SI}. 
c) Simulation snapshots corresponding to the velocities [I, II, III and IV] indicated in (a); color code as in Fig. {\ref{fig:MassInfoSpread}}.
}
\label{fig:pan4}
\end{figure}
%
% as we can see by the presence of large white islands.  

{\it Phase transition and criticality.--} 
Since the 3-state SIR model in 2D exhibits critical behavior~\cite{grassberger1983critical}, 
we investigate whether the 2-state active model in 2D  also displays criticality.  
In 2D,  agents move according to  $\dot{\mathbf{x}}_i = \text{V}[q_i] \hat{x}$, while transitions are controlled as before by Eq.~(\ref{eq:reaction}).  
The initial condition corresponds to a  polarized group of inactive agents, which are arranged equally spaced on the half-plane extending to the right. At $t=0$ there is either  
one active agent or the entire left boundary is active.  
Importantly, the reported results are not affected by the presence of polarization fluctuations,  if an alignment mechanism as in the Vicsek model is included in the model: particles quickly return to the polarization of the majority, see \cite{SI} for details and video S6.
%We also investigated a model where active agents perform Vicsek-like flocking dynamics obtaining qualitatively the same results
%
%
% Moreover, there exists a fundamental difference:  in the 2-state active model, it is not possible to observe firewalls of agents in state $\mathcal{R}$ that prevent islands of $\mathcal{S}$ agents to get infected as occurs in the SIR model. 

We observe that in  2D
the order parameter $\phi$ as function of  $v$ indicates the presence of a phase transition at a  critical speed $v_c$, with $v_c<0$;   Fig.~\ref{fig:pan4}(a). 
%
% For $v<v_{1}^*$, the propagation of the active state does not occur, while for $v>v_{2}^*$ an infinite propagation is possible {\color{blue}(Fig.~\ref{fig:pan4}(c) I and II)}. 
%
Interestingly, we observe that at $v \sim v_c$  there exists a remarkably large variability of possible outcomes for the same initial conditions; see video S7~\cite{SI} and Fig.~\ref{fig:pan4}c III and IV.
Specifically, the distribution of the number $s$ of active agents at $t\!\to\!\infty$,  $P(s)$, is power-law distributed, which is indicative of critical behavior. Specifically,  $P(s)\propto s^{-\beta}$, with $\beta \simeq -1$, and thus we observe from small to  giant, system spanning,  activity avalanches; Fig.~\ref{fig:pan4}(b). 
Compare these findings with what is known for  
Dynamic Percolation~\cite{munoz1999avalanche}. 
Criticality in our system is also manifested through the behaviour of susceptibility $\chi(\phi) = N(\langle \phi^2 \rangle - \langle \phi \rangle^2)$, which increases with system size alongside with a narrowing tendency (inset of Fig.~\ref{fig:pan4}(a)). Furthermore, we find that 
at $v \sim v_c$ 
the absence of characteristic correlation length, signifying that all length scales are possible~\cite{SI}. 
%
\iffalse
{\color{red} This behavior translates into a spatio-temporal dynamics in which 
the flow rate $Q(t)$ of active agents crossing at $x=0$ to the left half-plane 
fluctuates over time around a well-defined mean value  for small enough values of $|v|$, while it displays  large temporal fluctuations before the system falls into the absorbing phase;  Fig.~\ref{fig:pan4}(c). 
%
Fluctuations are evident  by looking at snapshots of the system; upper panels in Fig.~\ref{fig:pan4}(c). 
% It is the formation of percolating empty areas (in $\hat{y}$) that interrupts  activity propagation and pushes the system to the absorbing phase.  
%
Note that the average $\langle Q \rangle$ -- performed over time, while the system is in the active state -- as a function of $v$ exhibits a maximum; Fig.~\ref{fig:pan4}(d).  
%
This maximum results from the competition of high flows related to large $|v|$ values and  larger temporal fluctuations in the flow.}
%
\fi

 Note that the mapping onto the two-dimensional SIR lattice model is no longer exact, but an approximation. Since in the SIR model at the critical point, we observe that as the infection propagation advances,  
island of susceptible agents are left behind. These islands are protected by a layer of recovered agents and will never be infected~\cite{SI}. 
In contrast, in the studied 2-state system, due to the displacement of active agents, islands of inactive agents can be activated later on by a passing active agents~\cite{SI}.   
%
\iffalse
{\color{red}In the dynamical percolation transition in 2D, e.g.  SIR model~\cite{grassberger1983critical},  
critical behavior is observed only at the critical point. Here, however, we  do not need to fine-tune the value of $v$ to obtain a power-law distribution of cascade sizes; these are observed over a large range of $v$ values. 
%
This observation, together with the measured exponents of $p(s)$, leaves open the possibility of self-organized critical (SOC) behavior;  recall the well-established connection between SOC and systems with absorbing states~\cite{munoz1999avalanche}. 
%
Nevertheless, a detailed numerical investigation, exceeding our current computational capacity, and/or a renormalization group approach, beyond the scope of the current study, is required to settle this very interesting issue. }
\fi

In short, our study proves that a generic coupling between the agent's internal state and active movement has a profound impact 
on the physics of the propagation process, specifically: the coupling renders an active 2-state system into 
an effective 3-state (excitable) system that in 2D is able to display critical behavior. 
The obtained results are of key importance to the understanding of (i) criticality in active system, and of (ii)  how collectives distribute, process, and respond to the environmental information that is sensed by group members. 
The analysis also shows how the agent velocity selects the propagation regime.  
For example, when individuals move with velocity $0\!<\!v \! \ll \! c_0$, all group members get activated and move together, keeping the same (relative) group structure. %
On the other hand, if the first individual reacts to a threat, e.g. a predator, and chooses to move with velocity $v\gg c_0$, the individual will manage to 
seek cover  inside  the group, leaving a number of conspecifics  between its position and the predator, modifying the group structure, cf. Hamilton's selfish herd hypothesis~\cite{hamilton1971geometry}. 
%
%% We note that such a selfish behavior can only take place if individuals  move at a speed comparable to the one the activation travels: 
% Altering the relative position of individuals within the group requires high-speed values. 
%
Finally, the reported  predictions can be tested in experiments. In particular, preliminary results \cite{parisa2024} show that active 2-state models, in the spirit of the one analyzed here, reproduce collective diving events in fish~\cite{gomez2023fish}, and density and velocity waves in human crowds~\cite{bain2019dynamic}. These models are anticipated to also explain density fluctuations in sheep herds~\cite{gomez2022intermittent}, and activation waves in subcritical Quincke rollers~\cite{liu2021activity}. 

\begin{acknowledgments}
H.G. was supported by a PhD
grant of the French Ministry of Superior Education and Research. P.R. and F.P. acknowledge financial support from C.Y. Initiative of Excellence (grant Investissements d'Avenir ANR-16-IDEX-0008), INEX 2021 Ambition Project CollInt and Labex MME-DII, projects 2021-258 and 2021-297.
\end{acknowledgments}
\bibliographystyle{apsrev}
%\bibliography{biblio}

\end{document}